\documentclass[twocolumn,tighten,times]{aastex62}

\usepackage{empheq}

\newcommand\ergs{erg~s$^{-1}$}

\accepted{December 7, 2018}
\submitjournal{ApJ}
\shorttitle{Optically thick winds driven by supercritical accretion}
\shortauthors{Zhou et al.}

\begin{document}

\title{Evidence for Optically Thick, Eddington-limited Winds Driven by Supercritical Accretion}

\correspondingauthor{Hua Feng}
\email{hfeng@tsinghua.edu.cn}

\author{Yu Zhou}
\affiliation{Department of Engineering Physics and Center for Astrophysics, Tsinghua University, Beijing 100084, China}

\author{Hua Feng}
\affiliation{Department of Engineering Physics and Center for Astrophysics, Tsinghua University, Beijing 100084, China}

\author{Luis C.\ Ho}
\affiliation{Department of Astronomy, Peking University, Beijing 100087, China}
\affiliation{Kavli Institute for Astronomy and Astrophysics, Peking University, Beijing 100087, China}

\author{Yuhan Yao}
\affiliation{Cahill Center for Astrophysics, California Institute of Technology, MC 249-17, 1200 E California Boulevard, Pasadena, CA 91125, USA}

\begin{abstract}

Supercritical accretion onto compact objects powers a massive wind that is optically thick and Eddington-limited.  If most of the hard X-rays from the central disk are obscured by the wind, the source will display a blackbody-like spectrum with a luminosity scaled with the mass of the compact object.  From the \textit{Chandra} archive of nearby galaxies, we selected a sample of luminous and very soft sources and excluded contaminations from foreground objects and supernova remnants.  They are found to be preferentially associated with late-type galaxies. The majority of sources in our sample are either too hot or too luminous to be explained by nuclear burning on the surface of white dwarfs, and are argued to be powered by accretion.  The most likely explanation is that they are due to emission from the photosphere of a wind driven by supercritical accretion onto compact objects.  Their blackbody luminosity ranges from $\sim$$10^{37}$ to nearly $10^{40}$~\ergs,  indicative of the presence of both neutron stars and stellar-mass black holes. The blackbody luminosity also shows a possible bimodal distribution, albeit at low significance, peaked around the Eddington limit for neutron stars and stellar-mass black holes, respectively. If this can be confirmed, it will be smoking gun evidence that supercritical accretion powers thick winds.  Based on a wind model, the inferred mass accretion rate of these objects is around a few hundred times the Eddington rate, suggesting that they may be intermediate between the canonical ultraluminous X-ray sources and SS~433 in terms of the accretion rate.  

\end{abstract}


\section{Introduction}

The physics for accretion is still unclear in the case when the mass accretion rate is significantly above the critical value, the rate needed to power the Eddington luminosity. In the standard disk model \citep{Shakura1973}, supercritical accretion is not allowed in order to be self-consistent with the thin disk scenario.  The slim disk model \citep{Abramowicz1988} allows for an accretion rate moderately above the critical value.  This is because the radiation can be trapped in the accretion flow and swallowed by the central black hole\footnote{This is not valid if the accretor has a hard surface, such as a neutron star.} before it can diffuse to the surface of the disk \citep{Kato2008}, so that the local Eddington limit remains unbroken.  For the case with a highly super-Eddington accretion rate, the Polish doughnut model \citep{Abramowicz1978,Wielgus2016}, which assumes perfect fluid in gravity with a specific angular momentum distribution, resembles a thick accretion disk, or a fat torus, with a narrow funnel in the center. In these two super-Eddington models, no wind is necessary. 

On the other hand, due to the presence of strong radiation pressure, it seems inevitable that some or most of the accreting material will be launched into a wind in supercritical accretion \citep{Meier1982_PaperII,King2003,Poutanen2007,Shen2016}.  This has been confirmed by recent numerical simulations \citep{Ohsuga2011,Jiang2014,McKinney2015,Sadowski2016,Narayan2017,Dai2018}, which revealed two characteristics of the wind. First, the wind itself is optically thick, so that the emission generated from the disk covered by the wind will be thermalized and re-emitted as a blackbody-like spectrum at the photosphere of the wind. Second, in the central region, the optically thick wind encircles a low-density optically thin\footnote{The funnel is optically thin to absorption but thick to scattering \citep[\textit{cf.},][]{Jiang2014}.} funnel, where the hard X-rays from the central accretion disk can escape with some degree of mild, geometric beaming \citep[\textit{cf.},][]{Sadowski2016}. In addition, a turbulent wind may be another signature of supercritical accretion \citep{Takeuchi2013}.

Therefore, whether or not there exists a radiation-driven, optically thick wind may be a key to testing the supercritical accretion models.  \citet{Meier1982_PaperII} proposed a dynamical model for winds driven by supercritical accretion. He predicted that supercritical accretion onto compact objects will drive a wind that is optically thick and essentially spherical, with a luminosity close to the Eddington limit in a large fraction of the parameter space. In the wind model, some or most of the accreting materials will be launched into the wind and at least one Eddington rate is needed to be accreted and power the wind. 

To test the supercritical wind model in observations, ultraluminous X-ray sources \citep[ULXs; for a review see][]{Kaaret2017} are among the best targets.  Blueshifted absorption lines have been detected in bright ULXs \citep{Walton2016,Pinto2016,Pinto2017,Kosec2018}, but those lines trace the fast, optically thin part of the wind. The ``soft excess'' seen in the X-ray spectrum of ULXs can be modeled with a blackbody-like component and was argued to arise from the photosphere of the wind \citep{Poutanen2007,Middleton2015,Weng2018}. However, as the soft excess is usually not the dominant (\textit{e.g.}, $>$ 70\%) component in the spectrum and its characterization relies on a precise determination of the hard component and the absorption, its interpretation is often not unique \citep{Feng2009,Kajava2009,Miller2013}. 

Supersoft ULXs \citep[also called ultraluminous supersoft sources;][]{Liu2008} appear to be better targets for such a test, as the hard X-ray emission likely originated from the central funnel is secondary and, to some extent, does not influence the measurement of the soft X-ray emission from the wind photosphere. The energy spectrum for supersoft ULXs is dominated by a cool blackbody component \citep[\textit{e.g.},][]{Jin2011}, varying in a pattern close to $R_{\rm bb} \propto T_{\rm bb}^{-2}$ (or a constant $L_{\rm bb}$) that can be well explained by the optically thick wind model \citep{Shen2015,Soria2016,Urquhart2016,Feng2016}.\footnote{We note that alternative explanations exist for the soft blackbody emission, such as the thick accretion disk model \citep{Gu2016}.} 

In recent years, more and more ULXs were found to harbor a neutron star \citep{Bachetti2014, Fuerst2016, Israel2017,Israel2017a,Carpano2018}, whose Eddington limit is below the luminosity threshold for ULXs.  In addition, the outflow driven by supercritical accretion \citep{Meier1982_PaperII,Urquhart2016,Feng2016} may have a temperature higher than that for the canonical supersoft sources (SSSs) defined by \citet{DiStefano2004}. Thus, instead of focusing on supersoft and ultraluminous sources, we expand our search to luminous ($L_{\rm bb} \gtrsim 10^{37}$~\ergs) and very soft\footnote{We note that \citet{DiStefano2004} defined very soft sources as those with a blackbody temperature less than 0.3~keV (or 0.35 keV in their following papers), while we have a slightly higher threshold in this paper.} ($T_{\rm bb} \lesssim 0.4$~keV) X-ray sources. 

Based on a {\it Chandra} archival catalog of nearby galaxies within 50~Mpc \citep{She2017}, we conducted a population study of luminous very soft X-ray sources in this paper. The nature of these sources is not unique \citep[for brief reviews see][]{Distefano2010,Distefano2010a}, but our major science goal is to identify a class of them that is driven by supercritical accretion onto compact objects. The paper is constructed as follows. The source selection and the catalog assembly are described in \S~\ref{sec:sample}, the data analysis and results are presented in \S~\ref{sec:analysis}, and the interpretation of the results is elaborated and discussed in \S~\ref{sec:discussion}, with a summary in \S~\ref{sec:summary}. 

\section{Sample}
\label{sec:sample}

\citet{She2017} collected a full list of \textit{Chandra} observations with the advanced CCD imaging spectrometer (ACIS) as of 2016 March targeted on nearby galaxies within 50 Mpc. Our goal is to select non-nuclear point-like X-ray sources that are luminous and very soft. In practice, the selection is solely based on the spectral hardness and was done in two steps, a preliminary selection on the basis of the hardness ratio (HR),  and a further fine selection upon the energy spectrum.  The cut on the luminosity is not crucial, because most X-ray sources in external galaxies that can be detected with \textit{Chandra} in a reasonable exposure are close to or above our luminosity threshold.  We note that \citet{She2017} adopted distances from NED as they dealt with a large sample, while here we have looked into individual distances and chosen those that we think are most reliable. That is why the distances are not identical to those in \citet{She2017}, and some may exceed 50 Mpc.

\subsection{Selection based on HR}

The detection of off-nuclear X-ray sources was not reported in \citet{She2017}, who focused on the active galactic nuclei only. As a byproduct, off-nuclear X-ray sources on the same CCD with the center of the galaxy were detected as well along with the search for active nuclei.  For every source, the HR is calculated as the count ratio in two bands, $H/S$, where $H$ is the number of detected photons in the hard band ($2-8$ keV) and $S$ is that in the soft band ($0.3-2$ keV). 

The blackbody temperature is usually less than $\sim$0.2~keV for the thermal component in supersoft ULXs \citep{Urquhart2016}, and less than $\sim$0.4~keV for the soft excess in standard ULXs \citep{Feng2005,Feng2009,Kajava2009}. Therefore, we assume that the targets of our interest would have a spectrum softer than a blackbody of 0.4~keV.  Using the \textit{Chandra} PIMMS tool\footnote{http://cxc.harvard.edu/toolkit/pimms.jsp} and assuming a blackbody spectrum of 0.4~keV modified by interstellar absorption with a column density of $10^{21}$~cm$^{-2}$ (typical for sources in our final sample), an upper limit on the HR can be obtained as HR$_{\rm up}$ = 0.21. For accreting compact objects, their spectra may contain an additional power-law component even if the thermal component is dominant \citep[\textit{e.g.,}][]{Jin2011}. This may produce an overall spectrum deviating from a pure blackbody. If we assume a power-law model with a photon index of 5 \citep[this is empirically softer than the spectra of any Galactic X-ray binaries; see][]{Remillard2006} and the same interstellar absorption, the HR is about 0.02. Thus, we adopt 0.21 as a conservative criterion for the preliminary selection. 

Assuming Poisson fluctuation, the probability density of the intrinsic hardness ratio (HR$_0$) can be inferred using the Bayesian method \citep{Jin2006} as
\begin{equation}
p\left( {\rm HR}_0 = z \right) = \frac{z^H \times (H+S+1)\,!}{(z+1)^{H+S+2}H\,! \, S\,!} \; ,
\end{equation}
given the measured $H$ and $S$. From the probability distribution, we calculated the 90\% upper limit of the intrinsic hardness ratio (HR$_{90}$), and identify candidate very soft sources if ${\rm HR}_{90} < {\rm HR}_{\rm up}$. 
 
 \subsection{Selection based on spectra}
 
For the very soft candidates selected above, we then made a further selection based on their energy spectra, if they contain sufficient counts. With CIAO 4.9 and CALDB 4.7.6, new level-2 events files were created using the {\tt chandra\_repro} script. The task {\tt wavdetect} was used to detect sources and determine their source apertures, a $3\sigma$ elliptical region, while the background aperture was chosen from a nearby source-free region. The script {\tt specextract} was used to extract the source and background spectra and create response matrices. The spectra were grouped to have at least 15 counts per bin  in $0.3-8$ keV and fitted in XSPEC 12.9 if they have at least 100 counts in the same energy range. 

We tried to fit an absorbed power-law model ({\tt wabs * wabs * powerlaw}) to each spectrum to determine its apparent spectral hardness. The column density of the first absorption component was fixed at the Galactic value along the line-of-sight \citep{Kalberla2005}, while the second is a free component in the fit to account for absorption in the host galaxy or intrinsic to the source.  For any source if the derived power-law photon index is less than 5, we removed it from our sample. About 61\% of the sources selected by HR were discarded in this step. The remaining sources with an apparently soft spectrum constitute our final sample (Table~\ref{tab:sample}).

\subsection{Excluding non-accretion-powered objects}

Non-accretion-powered very soft sources include mostly foreground objects such as nuclear burning white dwarfs\footnote{To clarify, we note that their emission is not powered by accretion, although they are accreting objects.}  \citep{DiStefano2004} and X-ray active stars \citep{Covey2008}, and also supernova remnants \citep[SNRs; {\it e.g.},][]{2014ApJ...788...55B} in the host galaxies. They could appear bright and very soft in the X-ray band but are beyond the scope of this study. Foreground objects can be identified via their bright optical emission. For every object in our sample, we found their $B$ magnitude in optical catalogs, if available; see column 8 in Table~\ref{tab:sample}. Most were adopted from the USNO-B1.0 catalog \citep{Monet2003}; for sources 18, 66, and 14, the $B$ magnitude was found from the FONAC \citep{Kislyuk2000}, SPM 4.0 \citep{Girard2011}, and NOMAD \citep{Zacharias2004} catalog, respectively. We then searched the optical counterpart for every X-ray object in the GAIA Data Release 2 (DR2) catalog \citep{GaiaCollaboration2018} and quoted the parallax distance if there is a 3$\sigma$ measurement. The screening for non-accretion-powered objects was executed as follows, with remarks for some individual objects.

\begin{itemize}
\renewcommand{\labelitemi}{$-$}

\item Objects that have a positive parallax measurement with GAIA are marked as foreground stars, confirmed with a visual inspection of the optical image. This includes 46 objects.

\item Source 57 is not in the GAIA DR2 catalog, but has been identified as a high-proper motion star \citep{Wroblewski1991}. 

\item Source 30 and 94 are in the GAIA DR2 catalog but without a determination of the parallax or proper motion. There were low significance measurements in the Tycho-2 catalog \citep{Hog2000}. Source 30 is on the outskirts of NGC 1729. Source 94 is on the outer spiral arm of NGC 6946. Their blue magnitudes are around 11--12. We thus tend to identify them as possible foreground stars. They do not have high-quality data for spectral analysis and thus the exclusion of them has no influence on our conclusions.

\item Seven sources have been reported in the literature as candidate SNRs in their host galaxies. The references are listed in Table~\ref{tab:sample}.

\item Source 10 is identified as a Wolf-Rayet star in IC 1613 \citep{Armandroff1985}, but whether or not it is an accretion-powered source is unknown. Again, this object does not have data good enough for detailed spectral analysis. 

\item We note that source 65 is a known ULX, IXO 75 in NGC 5128 \citep{Colbert2002}, but GAIA observations nailed it down as a foreground object. A recent {\it Hubble} Space Telescope (HST) observation (dataset j9b408nrq) revealed that it is associated with a triple stellar system. 

\end{itemize}

We also calculated the distance of the X-ray source to the galactic center normalized to the optical isophotal radius along the direction of the X-ray source with respect to the center (see Table~\ref{tab:sample}, where $r_{25} = 1$ means the X-ray source is on the $D_{25}$ ellipse). There are two objects whose $D_{25}$ information is not available in RC3, but both are classified as foreground objects. All of the candidate objects of this study are within the $D_{25}$ ellipse of the galaxy. 

\begin{figure}[tb]
\centering
\includegraphics[width=\columnwidth]{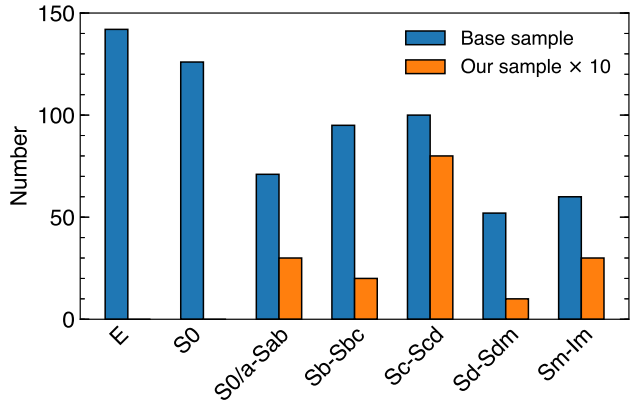}
\caption{Distribution of the host galaxy type for the input base sample (blue), from which we conducted the search, and our sample (orange; scaled by a factor of 10 for clarity). 
\label{fig:host}}
\end{figure}

To summarize, the whole sample contains 96 very soft X-ray sources selected from a \textit{Chandra} survey of nearby galaxies. Among them, 47 are found to be foreground objects, 7 are candidate SNRs in the host galaxy, and 2 are likely foreground or background objects. The remaining 40 objects are assumed to belong to the host galaxy and will be used for following study.  We note that the candidate sources have faint optical emission on HST images, if available, which strengthens their association with the host galaxy (this will be reported in a companion paper). However, we note that there is no solid evidence for every individual source to be firmly associated with the host galaxy. The Hubble type distribution of the host galaxies of these 40 objects are shown in Figure~\ref{fig:host}.  As one can see, the very soft sources are preferentially associated with late-type galaxies. 

For these 40 objects, we collected their \textit{Chandra} observations and listed necessary information in Table~\ref{tab:obs}, including the exposure time, source counts in several energy bands,  and the hardness ratios as defined above. We also checked the temporal variability for each source using the Kolmogorov-Smirnov (KS) test against a constant flux in all available \textit{Chandra} observations, and quoted the significance against a constant flux.  

\section{Spectral analysis and results}
\label{sec:analysis}

\begin{figure*}[ht]
\centering
\includegraphics[width=0.8\textwidth]{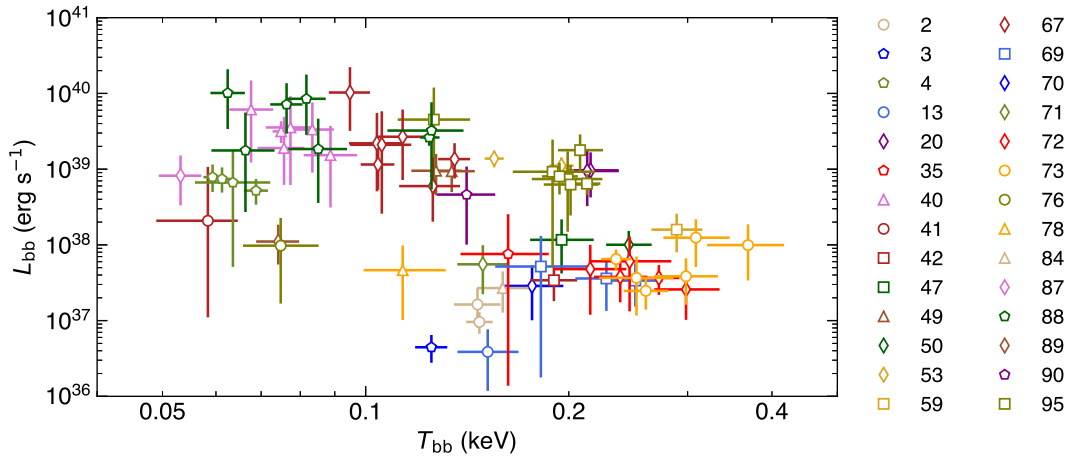}
\caption{Blackbody luminosity versus temperature for observations in our sample with at least 100 counts in 0.3--8 keV. Observations with a non-detection in the blackbody luminosity are not shown. Multiple observations for the same object are shown with the same symbol.  The errors are of 1$\sigma$.
\label{fig:data}}
\end{figure*}

\begin{figure}[ht]
\includegraphics[width=\columnwidth]{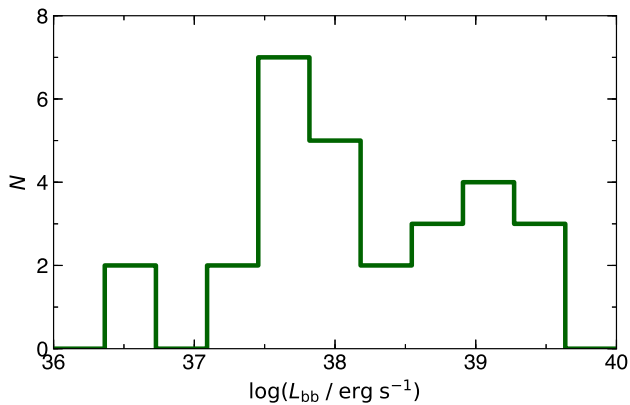}
\caption{Distribution of the intrinsic blackbody luminosity for objects in our sample. For objects with multiple observations, the median luminosity was chosen (or the mean of the central two values if the number is even). There seems to be two peaks in the distribution, one slightly below the Eddington limit for neutron stars ($10^{38}$~\ergs) and the other around the Eddington limit for stellar-mass black holes ($10^{39}$~\ergs). 
\label{fig:lumdist}}
\end{figure}

For sources with at least 100 photons detected in 0.3--8 keV, we fitted their spectra with physical models. We first attempted to fit the source spectrum with a single blackbody model subject to interstellar absorption.  As mentioned above, the absorption is implemented with two components, one for Galactic and the other for extragalactic. In some cases, if the extragalactic absorption needs to be zero to find the minimum, it is removed and only the fixed Galactic absorption is used.  If the single blackbody model does not provide an adequate fit (usually in the high-energy band), resulting in a null hypothesis probability less than 0.05, we then tried to add an additional power-law component to account for the high-energy emission above $\sim$1.5~keV.  If the power-law component is weak and hard to be determined, dominant only in 1 or 2 spectral bins, we discard the few bins in the hard tail and use a single blackbody to find an adequate fit.  With such a treatment, the effect on $T_{\rm bb}$ and $L_{\rm bb}$ is about $\sim$10\%, much smaller than the statistical errors, evaluated using spectra where the power-law component can be determined. For bright sources, we estimated the pileup fraction following the \textit{Chandra} ABC Guide to Pileup.\footnote{http://cxc.harvard.edu/ciao/download/doc/pileup\_abc.pdf}  Fortunately, most of our sources are not extremely bright, so that the {\tt pileup} model in XSPEC can be used to correct for the moderate pileup effect, except for spectra from three observations (ObsIDs 310, 390, and 402, respectively for sources 4, 40, and 51), which were not used.  The pileup fraction is found to be 0.10, 0.03, 0.03, 0.11, and 0.06, respectively, for ObsID 309, 735, 1575,  1854, and 4737. The spectral parameters with 1$\sigma$ error bounds are listed in Table~\ref{tab:spec}, in which we are most interested in the blackbody temperature $T_{\rm bb}$ and blackbody luminosity $L_{\rm bb} \equiv 4 \pi R_{\rm bb}^2 \sigma T_{\rm bb}^4$.  We note that, depending on the temperature, the X-ray luminosity is only a fraction of the bolometric blackbody luminosity.  

As the sensitive energy band of \textit{Chandra} cannot go below 0.3~keV (or 0.5~keV in some cases), the peak of the energy spectrum for blackbody emission with a temperature below 0.1~keV cannot be covered by the observation. This may introduce a degeneracy between the absorption column density and the blackbody temperature; less absorbed, hotter objects may appear like a more absorbed cooler object.  We ran simulations to test whether or not the derived blackbody luminosities and temperatures are reliable. We adopted $T_{\rm bb}$ and $L_{\rm bb}$ at the extremes of our parameter space, simulated 10,000 mock spectra for each combination with the number of photons similar to that observed, and fitted them following the same protocol as above. The derived $T_{\rm bb}$ and $L_{\rm bb}$ from the mock spectra have a mean consistent with the input value, suggestive of an unbiased measurement; the distribution width is comparable to or smaller than that from real measurements,  indicating that the uncertainty range calculated by the {\tt error} command in XSPEC is a safe estimate. We also expanded the test to regimes below the lowest temperature that we measured, and found that the temperature for sources in that parameter space ($T_{\rm bb} = 0.03-0.05$~keV) will not be systematically overestimated. This justifies that the derived blackbody luminosities and temperatures are unbiased and the errors are not underestimated. 

We compared our results with those reported in the literature. If identical models were adopted, well consistent results were obtained as well \citep[{\it e.g.},][]{Urquhart2016}. For the brightest sources, even the two-component model, blackbody plus power law, is unable to fit the spectrum adequately, and soft residuals can be seen and phenomenologically fitted with thermal plasmas and/or absorption edges \citep[{\it e.g.},][]{Jin2011,Urquhart2016}. \citet{Pinto2017} pointed out that the residuals can be explained by the fast wind model, by comparing the spectra obtained from the gratings and from the CCDs. Study of the soft residuals is beyond the scope of this study.  We argue that the derived properties of the blackbody component are not strongly affected if one ignores the residuals in the spectral fitting for the high-quality data. For example, the spectrum that contains the largest number of photons in our sample is from ObsID 934 for M101 ULX-1 (source 88); the complex model \citep{Soria2016} and our simple, two-component model give a difference of 8\% for $T_{\rm bb}$ and 40\% for $L_{\rm bb}$, acceptable for our following physical interpretation. 

For sources with a positive detection in the blackbody luminosity ($L_{\rm bb}$ larger than its 1$\sigma$ error), the $L_{\rm bb}$ versus $T_{\rm bb}$ diagram is shown in Figure~\ref{fig:data}.  The upper limits from observations with a non-detection are loose and thus not shown. Multiple observations from the same object are indicated with the same symbol. The histogram for $\log L_{\rm bb}$ is displayed in Figure~\ref{fig:lumdist}, on which the median luminosity, if there is more than one observation, was chosen for each source. 

\section{Discussion}
\label{sec:discussion}

Here we discuss the physical nature of the sources in our sample. We will show below that the majority of objects in the sample are powered by accretion, and the most likely explanation is that they are hyper-accreting objects where emission from an optically thick wind dominates the observed X-ray spectrum.

\subsection{Unlikely explanations: white dwarfs and intermediate-mass black holes}

Canonical SSSs are thought to be powered by steady nuclear burning on the surface of an accreting white dwarf in a close binary system \citep{vandenHeuvel1992}. They may have a bolometric luminosity up to $10^{38}$~\ergs\ with a temperature below $\sim$80~eV \citep{Greiner2000}. In our sample, only a couple sources may occupy this regime in the $L_{\rm bb} - T_{\rm bb}$ diagram but the majority do not. The same conclusion can be drawn for supersoft X-ray novae, which are another major population of canonical SSSs \citep{Orio2010}.  We note that these sources are naturally excluded in this study due to selection effect. 

Subcritically accreting X-ray binaries may be soft and show a thermal spectrum in the thermal dominant state \citep{Remillard2006}. However, to explain the observed luminosity and temperature for objects in our sample, it requires intermediate-mass black holes (IMBHs).  Although we cannot rule out such a possibility, the study of ULXs has pointed out that an IMBH interpretation is not needed for most of them. Also, X-ray binaries in the thermal dominant state are thought to have an accretion disk extending all the way to the innermost stable circular orbit, and display a disk luminosity that is varying in proportion to the fourth power of the disk inner temperature. However, none of our sources with multiple observations shows such a behavior. 

A collateral evidence for accretion is that, for sources in the sample with at least 1000 counts in any observation, 4 out of 5 show short-term variability ($>3\sigma$ with KS test), while the fraction is only 1/9 of foreground objects. Thus, we conclude that the sources in the sample are accreting compact objects but not in the regime of accretion rate that traditional X-ray binaries occupy.  The most likely explanation is that they are powered by supercritical accretion. We hereby discuss the results in the frame of such a model. 

\subsection{The wind interpretation}

For a quantitative comparison with the results, we adopt the 1D radiation hydrodynamic wind model depicted by \citet{Meier1982_PaperII,Meier1982_PaperIII,Meier1982_PaperIV}, and also in his recent book \citep{Meier2012}. In the case of supercritical accretion onto stellar-mass compact objects, the wind will be optically thick in the beginning and the radiation luminosity will keep constant if one assumes local thermodynamic equilibrium in the wind.  That means, if the input radiation at the base of the wind is Eddington-limited, so is the reprocessed radiation at the photosphere of the wind.

As there is no self-consistent model taking into account both the supercritical inflow and outflow, one of the major challenges is to define the inner boundary conditions of the wind, {\it i.e.}, at which radii the wind is launched, with what density, velocity, pressure, and temperature. A good approximation, to first order, is to assume that the wind launches at the point where the local Eddington limit is approached, or the so-called spherization radius \citep{Shakura1973,Poutanen2007}. Here, we follow the recipe in \citet{Meier2012} and assume that the wind launches above a slim disk at the radius close to the point where the advective loss is comparable to the radiative cooling. This is also the radius where the total luminosity approaches the Eddington limit. In contrast to the disk model, most of the accreting material is converted into the wind with the same equation of state. As accretion at one Eddington rate is still needed to power the wind, this is valid only if $\dot{m} \gg 1$. More detailed explanations of the model and the final recipes used here are elaborated in Appendix~\ref{sec:model}.

\begin{figure}[tb]
\includegraphics[width=\columnwidth]{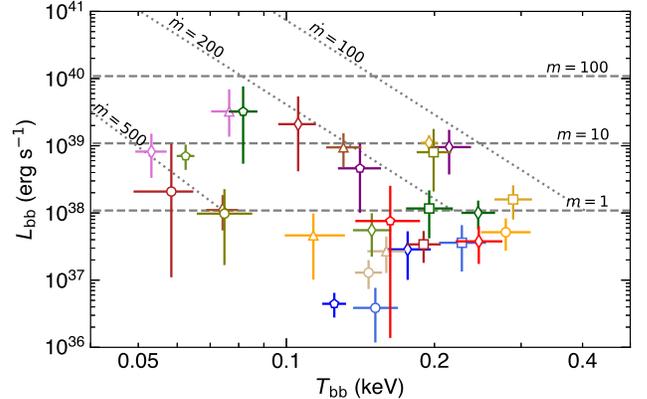}
\caption{Blackbody luminosity versus temperature for objects in our sample superposed with model predictions. The data are the same as in Figure~\ref{fig:data}, but for objects with multiple observations, only the median value is used, \textit{i.e.}, Figure~\ref{fig:lumdist} is the histogram of the data in this plot. The horizontal dashed lines are the predicted $L_{\rm bb}$ versus $T_{\rm bb}$ relation given the same compact object mass. The dotted lines indicate the $L_{\rm bb}$ versus $T_{\rm bb}$ relation for compact objects of different masses but at the same mass accretion rate. 
\label{fig:data_model}}
\end{figure}

In Figure~\ref{fig:data_model}, we plotted the model-predicted wind luminosity and temperature given the compact object mass and accretion rate.  As one can see, it suggests that objects in our sample may cluster at two regimes, one with a compact object mass close to 1~$M_\sun$ (neutron stars), and the other with a mass around 10--100~$M_\sun$ (stellar-mass black holes). They share a similar accretion rate range, $\dot{m} = 100 - 500$.  The derived mass accretion rate is degenerate with the $\alpha$ value ($\dot{m} \propto \alpha^{2/5}$ for subsonic solutions and $\dot{m} \propto \alpha^{1/3}$ for supersonic solutions). Here we have assumed $\alpha = 0.1$, and a smaller $\alpha$ will lead to a lower accretion rate.  We also note that the Comptonization $y$ parameter at the photosphere may be large, so that moderate or weak Comptonization may exist beyond the photosphere, but is not corrected here. If we follow \citet{Meier1982_PaperIII} to correct for the Comptonization effect, the derived $\dot{m}$ range will become $\dot{m} = 70 - 500$ for our sample, because the $y$ parameter on the photosphere is large only when the accretion rate is relatively low. Thus, this correction was not employed in view of the fact that some other parameters and the boundary conditions may have uncertainties of the same order of magnitude.

As for the blackbody emission from the wind photosphere, it is reasonable to assume that it caps at the Eddington limit. In our case, we have $L_{\rm bb} = \frac{3}{4} L_{\rm Edd}$ simply due to the assumptions of the boundary conditions. If the wind is launched before the Eddington limit is approached, which is likely in reality, the wind luminosity may be lower than the Eddington limit. This may explain why some data points in Figure~\ref{fig:data_model} cluster at a region slightly below the Eddington luminosity for a neutron star.  

Again, we want to emphasize that this model may be accurate only to an order of magnitude. The wind model itself is accurate enough \citep{Meier1979_PaperI,Meier1982_PaperII} but valid only for the 1D case, while the real wind may be 2D.  More importantly, a global solution to the inflow and outflow is needed to get the boundary conditions for the wind.

\subsection{A possible bimodal feature}

The blackbody luminosity distribution seems to have two peaks (Figure~\ref{fig:lumdist}), one between $10^{37}$ to $10^{38}$~\ergs, and the other around $10^{39}$~\ergs.  This resembles the Eddington limit for neutron stars and stellar-mass black holes, respectively, and argues in favor of the above interpretation that these objects are indeed Eddington-limited outflows powered by supercritical accretion onto compact objects.  As there are only 28 objects in the histogram, the significance for a bimodal distribution is not significant from a statistical point of view.  We adopted the Gaussian mixture modeling test \citep{Muratov2010} to see whether or not the distribution is bimodal. The test for two Gaussians with the same width over one Gaussian results in a $p$-value of 0.07, while the test for two Gaussians with different widths gives a $p$-value of 0.2.  This is not surprising due to the relatively small sample size.  

Thus, the bimodal feature is marginally significant only if the two Gaussians have the same variance.  Plus, our sample is subject to a selection effect that faint and cool objects cannot be detected. With these caveats, one should be cautious about the reality of the apparent bimodality. If the feature is not real, it will not reject the above inference, which is still physically valid and self-consistent, and one still requires the presence of both neutron stars and stellar-mass black holes to explain the range of the blackbody luminosity.  If the bimodal feature is firmly observed in the future, it will be a smoking gun for the above interpretation. 

\subsection{Connection with ULXs and SS~433}

The association with late-type galaxies for objects in our sample is consistent with the properties of ULXs \citep{Swartz2011}. This leads to a speculation that they are high-mass X-ray binaries undergoing thermal timescale mass transfer \citep{King2001}, which can provide the extreme mass accretion rate needed to fit the observations.  Most of the often-studied supersoft ULXs \citep[see][]{Urquhart2016} are included in our sample,\footnote{Sources 5, 40, 49, 67, and 88, are respectively, the supersoft ULX in NGC 247, M81, NGC 4038/39, M51, and M101; those in NGC 300 and NGC 4631 are discarded as they have fewer than 100 counts in \textit{Chandra} observations.}  composing the sub-population at the high-luminosity end.  Compared with standard ULXs, which have substantial emission above 2~keV, the supersoft ULXs can also be explained under the ``unification model'' of ULXs as due to geometric effect \citep{Middleton2015,Urquhart2016}. The hard X-ray emission from the central disk is collimated by the funnel and is more or less beamed depending on the position of the scattersphere\footnote{Here we follow the terminology in \citet{Meier1982_PaperII}.} on top of the funnel.  They appear supersoft either because they are viewed at high inclination angles, or because they undergo a higher accretion rate that produces a narrower funnel.  At least, the supersoft ULX in NGC 247 is a system likely with a higher accretion rate compared with standard ULXs, as a transition between a standard soft regime and a supersoft regime was observed \citep{Feng2016}. Plus, albeit at low significance, its fast wind is found to be relatively slower than that from standard ULXs \citep{Pinto2017}, consistent with the wind model that the velocity inversely scales with the accretion rate ({\it cf.}, Equation \ref{eq:v0}). 

Considering that the photosphere temperature inversely scales with the accretion rate (Equations \ref{eq:T_star_sub} \& \ref{eq:T_star_sup}), and in view of the fact the blackbody temperatures in supersoft ULXs are systematically lower than those in standard ULXs, we argue that the softer systems are on average sustaining a higher accretion rate.  However, one should keep in mind the caveat that here we consider a 1D wind model, but in practice the wind is at least 2D and its photosphere temperature may be lower if viewed at higher inclinations because the wind materials from that part may originate from an outer part of the disk. 

An extreme case for such a wind would be the Galactic microquasar SS~433, whose mass accretion rate assuming a stellar-mass black hole is on the order of $10^3$~$\dot{M}_{\rm Edd}$ \citep{Fabrika2004}. The temperature of the photosphere of its slow wind is $\gtrsim 5 \times 10^4$~K, peaked in the UV band, consistent with that predicted by the wind model.  

Therefore, a reasonable speculation is that standard ULXs are systems with $\dot{m} \approx 1 - 10^2$, where the hard X-rays from the central funnel can still illuminate to a considerable solid angle, and supersoft ULXs and very soft sources in our sample are systems with $\dot{m} \approx 10^2$, where the central funnel is rather narrow. Again, we emphasize that it cannot be ruled out that these very soft systems are high-inclination systems,  as the photosphere temperature and the velocity of the fast wind could be a function of the viewing angle. 

\section{Summary}
\label{sec:summary}

We collected a sample of luminous ($10^{37} - 10^{40}$~\ergs) and very soft (0.05--0.4~keV) sources in nearby galaxies, after removing contaminations from foreground objects and SNRs in the host galaxy.  Their emission is dominated by a blackbody component. The most likely interpretation is that they are powered by supercritical accretion onto compact objects, where the blackbody emission arises from the photosphere of an optically thick wind, which is Eddington-limited.  To explain the observed range of the blackbody luminosity, both neutron stars and stellar-mass black holes are required as the accretor. This is in line with the fact that a considerable fraction of ULXs may contain neutron stars \citep{Middleton2017} but they cannot be easily detected with current telescopes \citep{Pintore2017}.  A possible bimodal feature in the distribution of the blackbody luminosity is seen, but at low significance.  The two apparent peaks are located just around the Eddington limits for neutron stars and stellar-mass black holes. Applying the wind model by \citet{Meier1982_PaperII}, the inferred accretion rate is in the range of $\dot{m} = 100-500$.  This depends on the assumptions of the boundary conditions (where the wind is launched and its equation of state is set) and some parameters ({\it e.g.}, $\alpha$), but is considered correct as an order-of-magnitude estimate.  These objects are cousins of ULXs but with the majority of hard X-rays obscured by the thick wind.  They seem to be ULXs with an even higher accretion rate, but the scenario that they are standard ULXs viewed at high inclinations cannot be ruled out.

\acknowledgments We thank the referee for useful comments, and are grateful to David Meier and Weimin Gu for insightful discussions and comments. HF acknowledges funding support from the National Key R\&D Project (grants Nos.\ 2018YFA0404502 \& 2016YFA040080X), and the National Natural Science Foundation of China under the grant Nos.\ 11633003 \& 11821303.  
LCH was supported by the National Key R\&D Program of China (grant No.\ 2016YFA0400702) and the National Science Foundation of China under grant Nos.\ 11473002 and 11721303.




\startlongtable
\begin{deluxetable*}{lccllcccll}
\tablewidth{0pc}
\tabletypesize{\scriptsize}
\tablecaption{Very soft sources along the line-of-sight of nearby galaxies \label{tab:sample}}
\tablehead{
\colhead{ID} & \colhead{R.A.} & \colhead{Decl.} & \colhead{Galaxy} & \colhead{$d_{\rm galaxy}$} & \colhead{Refs.} & \colhead{$r_{25}$} & \colhead{$d_{\rm parallax}$} & \colhead{$B$} & \colhead{Remark} \\
\colhead{} & \colhead{(J2000)} & \colhead{(J2000)} & \colhead{} & \colhead{(Mpc)} & \colhead{} & \colhead{} & \colhead{(kpc)} & \colhead{(mag)} & \colhead{}\\
\colhead{(1)} & \colhead{(2)} & \colhead{(3)} & \colhead{(4)} & \colhead{(5)} & \colhead{(6)} & \colhead{(7)} & \colhead{(8)} & \colhead{(9)} & \colhead{(10)}
}
\startdata

    1               &   00:14:45.72     &   $-$39:14:35.6   &   NGC 55          &   2.1             &   21              &   0.57            &   0.19            &   13.1   &   FS                       \\
    2               &   00:42:43.30     &   +41:13:19.4     &   NGC 244 (M31)   &   0.78            &   3               &   0.03            &                   &      &                                \\
    3               &   00:42:47.87     &   +41:15:49.8     &   NGC 244 (M31)   &   0.78            &   3               &   0.01            &                   &      &                                \\
    4               &   00:42:52.52     &   +41:15:39.8     &   NGC 244 (M31)   &   0.78            &   3               &   0.03            &                   &      &                                \\
    5               &   00:47:03.88     &   $-$20:47:44.0   &   NGC 247         &   3.4             &   14              &   0.22            &                   &      &                                \\
    6               &   00:47:12.00     &   $-$25:20:37.8   &   NGC 253         &   3.56            &   24              &   0.24            &                   &      &                                \\
    7               &   00:47:32.98     &   $-$25:18:46.3   &   NGC 253         &   3.56            &   24              &   0.17            &                   &      &                                \\
    8               &   00:52:14.71     &   +47:37:01.5     &   NGC 278         &   12.1            &   11              &   2.27            &   0.30            &   11.5   &   FS                       \\
\enddata

\tablecomments{
Col.~(1): Source ID.
Col.~(2): Right ascension of the X-ray position.
Col.~(3): Declination of the X-ray position.
Col.~(4): Name of the host galaxy or that along the line of sight.
Col.~(5): Distance to the galaxy.
Col.~(6): Reference for the distance to the galaxy.
Col.~(7): Distance to the galaxy center normalized to the isophotal radius where the blue surface brightness is 25~mag~arcsec$^{-2}$.
Col.~(8): Parallax distance in GAIA DR2 for the optical counterpart.
Col.~(9): $B$ magnitude for the optical counterpart.
Col.~(10): Possible nature for the source: FS -- foreground star, WR -- Wolf-Rayet star in the host galaxy, SNR -- supernova remnant in the host galaxy with reference, FS -- foreground star.
}

\tablerefs{(1) \citet{1988ngc..book.....T}, (2) \citet{1989ApJS...69..763F}, (3) \citet{1998AJ....115.1916H}, (4) \citet{2000AA...362..544K}, (5) \citet{2000ApJ...529..745F}, (6) \citet{2001ApJ...546..681T}, (7) \citet{2002AJ....123.1389B}, (8) \citet{2002Ap.....45..253T}, (9) \citet{2003AA...410..445M}, (10) \citet{2004ApJ...608...42S}, (11) \citet{2006ApJS..164...81M}, (12) \citet{2006ApJS..165..108S}, (13) \citet{2008AJ....136.1482S}, (14) \citet{2009ApJ...700.1141G}, (15) \citet{2009ApJ...704.1586S}, (16) \citet{2009ApJS..183...67D}, (17) \citet{2010ApJ...724..657B}, (18) \citet{2010PASA...27..457H}, (19) \citet{2011AA...532A.104N}, (20) \citet{2011ApJ...733..124S}, (21) \citet{2011ApJ...738..150T}, (22) \citet{2011ApJ...743..176G}, (23) \citet{2012ApJ...758...11C}, (24) \citet{2013AJ....146...86T}, (25) \citet{2014ApJ...782...98B}, (26) \citet{2016AJ....152...50T}, (27) \citet{2016AJ....152..144M}, (28) \citet{2016ApJ...822...70L}, (29) \citet{2017ApJ...834..174K}, (30) \citet{DiStefano2004}, (31) \citet{2009ApJ...703..370C}, (32) \citet{2012AA...544A..81H}, (33) \citet{2012ApJS..203....8B}, (34) \citet{2014ApJ...788...55B}.}

\end{deluxetable*}

\startlongtable
\begin{deluxetable*}{llllllllll}
\tablewidth{0pc}
\tabletypesize{\scriptsize}
\tablecaption{\textit{Chandra} observations of very soft X-ray sources in nearby galaxies \label{tab:obs}}
\tablehead{
\colhead{ID} & \colhead{Host galaxy} & \colhead{ObsID} & \colhead{Exposure} & \colhead{Counts} & \colhead{Counts} & \colhead{Counts} & \colhead{HR} & \colhead{HR$_{90}$} & \colhead{$\sigma_{\rm KS}$} \\
\colhead{} & \colhead{} & \colhead{} & \colhead{(ks)} & \colhead{(0.3--1 keV)} & \colhead{(1--2 keV)} & \colhead{(2--8 keV)} & \colhead{} & \colhead{} & \colhead{} \\
\colhead{(1)} & \colhead{(2)} & \colhead{(3)} & \colhead{(4)} & \colhead{(5)} & \colhead{(6)} & \colhead{(7)} & \colhead{(8)} & \colhead{(9)} & \colhead{(10)}
}
\startdata

    2               &   NGC 244 (M31)   &   1575            &   37.7            &   335             &   123             &   9               &   0.019           &   0.028           &   0.3             \\
    2               &   NGC 244 (M31)   &   309             &   5.1             &   103             &   24              &   3               &   0.023           &   0.055           &   3.0             \\
    3               &   NGC 244 (M31)   &   1575            &   37.7            &   163             &   27              &   0               &   0               &   0.012           &   1.4             \\
    4               &   NGC 244 (M31)   &   1575            &   37.7            &   2477            &   36              &   1               &   0               &   0.001           &   2.2             \\
    4               &   NGC 244 (M31)   &   309             &   5.1             &   538             &   24              &   0               &   0               &   0.004           &   0.4             \\
    4               &   NGC 244 (M31)   &   1854            &   4.7             &   346             &   9               &   0               &   0               &   0.006           &   0.4             \\
    4               &   NGC 244 (M31)   &   310             &   5.1             &   334             &   4               &   0               &   0               &   0.006           &   0.5             \\
    4               &   NGC 244 (M31)   &   303             &   11.8            &   270             &   3               &   0               &   0               &   0.008           &   0.4             \\
\enddata

\tablecomments{ObsID is the \textit{Chandra} observation ID. HR is defined as the ratio of counts in 2--8 keV to those in 0.3--2 keV. HR$_{90}$ is the 90\% upper limit of the intrinsic hardness ratio inferred using the Bayesian method. $\sigma_{\rm KS}$ is the significance in units of sigma that the source deviates from constant flux. }

\end{deluxetable*}

\startlongtable
\begin{deluxetable*}{llllllllllll}
\tablewidth{0pc}
\tabletypesize{\scriptsize}
\tablecaption{Spectral parameters for sources in our sample \label{tab:spec}}
\tablehead{
\colhead{ID} & \colhead{ObsID} & \colhead{$N_{\rm H, Gal}$} & \colhead{$N_{\rm H, ext}$} & \colhead{Model} & \colhead{$T_{\rm bb}$} & \colhead{log($L_{\rm bb}$/erg $\rm s^{-1}$)} & \colhead{$\Gamma_{\rm PL}$} & \colhead{$A_{\rm PL}$} & \colhead{$\chi^2$} & \colhead{d.o.f.} \\
\colhead{} & \colhead{} & \colhead{($10^{21}$ $\rm cm^{-2}$)} & \colhead{($10^{21}$ $\rm cm^{-2}$)} & \colhead{} & \colhead{(keV)} & \colhead{} & \colhead{} & \colhead{($10^{-5}$)} & \colhead{} & \colhead{} \\
\colhead{(1)} & \colhead{(2)} & \colhead{(3)} & \colhead{(4)} & \colhead{(5)} & \colhead{(6)} & \colhead{(7)} & \colhead{(8)}  & \colhead{(9)} & \colhead{(10)} & \colhead{(11)} 
}
\startdata

    2               &   1575            &   1.25            &   \nodata                             &   BB              &   $0.147^{+0.006}_{-0.006}$           &   $36.98^{+0.11}_{-0.14}$             &                                       &                                       &   35.6            &   23              \\
    2               &   309             &   1.25            &   \nodata                             &   BB              &   $0.146^{+0.011}_{-0.011}$           &   $37.21^{+0.20}_{-0.29}$             &                                       &                                       &   2.41            &   5               \\
    3               &   1575            &   1.27            &   \nodata                             &   BB              &   $0.125^{+0.006}_{-0.006}$           &   $36.65^{+0.15}_{-0.19}$             &                                       &                                       &   4.75            &   9               \\
    4               &   1575            &   1.27            &   $0.34^{+0.14}_{-0.14}$              &   (pileup) BB     &   $0.059^{+0.002}_{-0.002}$           &   $38.89^{+0.15}_{-0.18}$             &                                       &                                       &   45.5            &   33              \\
    4               &   309             &   1.27            &   \nodata                             &   (pileup) BB     &   $0.069^{+0.003}_{-0.002}$           &   $38.72^{+0.14}_{-0.17}$             &                                       &                                       &   15.1            &   21              \\
    4               &   1854            &   1.27            &   $0.7^{+0.5}_{-0.5}$                 &   (pileup) BB     &   $0.052^{+0.005}_{-0.004}$           &   $39.5^{+0.8}_{-\infty}$             &                                       &                                       &   7.13            &   11              \\
\enddata

\tablecomments{$N_{\rm H, Gal}$ is the Galactic absorption along the line-of-sight to the source. $N_{\rm H, ext}$ is the absorption beyond the Milky Way. Model: BB stands for blackbody, PL for power-law, and pileup for the pileup model in XSPEC. $\Gamma_{\rm PL}$ is the power-law photon index. $A_{\rm PL}$ is the normalization of the power-law model at 1 keV, in units of photons~cm$^{-2}$~s$^{-1}$~keV$^{-1}$. Errors are quoted at 1~$\sigma$.}

\end{deluxetable*}

\clearpage


\appendix

\section{Meier's model for optically thick wind driven by supercritical accretion}
\label{sec:model}

This section is not original, but we include it to summarize the optically thick wind model driven by supercritical accretion proposed by \citet{Meier1982_PaperII,Meier1982_PaperIII}.  The readers can also refer to \S~13.1.2 in \citet{Meier2012}. All of the quantities used in this paper are defined in Table~\ref{tab:def}.  We note that solar abundance is assumed, but this may not be the case for ULXs \citep[\textit{e.g.,}][]{Zampieri2009}. The abundance has little effect on the results compared with other parameters.

\begin{deluxetable}{ll}[b]
\tablecolumns{2}
\tablewidth{0pc}
\tablecaption{Definitions of physical quantities \label{tab:def}}
\tablehead{
\colhead{Definition or assumption} & \colhead{Note}
}
\startdata
$m = M / M_\sun$  & dimensionless compact object mass \\
$L_{\rm Edd} = 4 \pi G M c / \kappa_{\rm es} = 1.5 \times 10^{38} m$~\ergs  & Eddington luminosity \\
$\epsilon_{\rm acc} = 0.1$  & accretion efficiency \\
$\alpha = 0.1$ & dimensionless viscosity parameter \\
$\dot{M}_{\rm Edd} = L_{\rm Edd} / \epsilon_{\rm acc} c^2$  & mass accretion rate needed to power the Eddington luminosity \\
$\dot{m} = \dot{M} / \dot{M}_{\rm Edd}$  & dimensionless mass accretion rate \\
$X = 0.71$, $Y = 0.27$, $Z = 0.02$  & metallicity assuming solar abundance \\
$\mu = 1 / (2X + 0.75Y + 0.56Z) = 0.61$  & average molecular weight assuming full ionization \\
$\kappa_{\rm es} = (0.2 \; {\rm cm^2 \; g^{-1}}) (1 + X) = 0.342$~cm$^2$~g$^{-1}$ & electron scattering opacity \\
$K_0 = 1.55 \times 10^{24}$~cm$^5$~g$^{-2}$~K$^{7/2}$ & Rosseland mean \\
$\kappa_{\rm ff/bf} = K_0 \rho T^{-7/2}$   & free-free and bound-free absorption opacity \\
\enddata
\end{deluxetable}

The base or the inner boundary of the wind is referred to as the injection region \citep{Meier1982_PaperII}, where the wind is launched and starts to accelerate. Here we adopt the slim disk solutions described in \citet[][see Equations 12.52 and 12.53]{Meier2012} and assume that the wind launches at the radius $r_{\rm i} = \dot{m} R_{\rm isco}$, where $R_{\rm isco}$ is the radius of the innermost stable circular orbit around a non-spinning black hole.  This injection radius is close to the place where advective and radiative cooling become comparable. The radiation pressure will remove excess material from the disk into the wind, and the remaining material ($\sim \dot{m}$) will go all the way onto the central compact object and produce a luminosity to power the wind. Inserting $r_{\rm i}$ to the slim disk solutions, one obtains the inner boundary conditions for the wind (radius, density, velocity, pressure, and temperature) at the injection region, which are valid only if $\dot{m} \gg 1$.
\begin{subequations}\label{eq:inj}
\begin{align}
r_{\rm i} &= (8.9 \times 10^5 \; {\rm cm}) \; m \dot{m} \; ,\\
\rho_{\rm i} &= (2.1 \times 10^{-5} \; {\rm g \; cm^{-3}}) \; m^{-1} \dot{m}^{-1/2} \alpha^{-1} \; ,\\
v_{\rm i} &= (9.2 \times 10^{9} \; {\rm cm \; s^{-1}}) \; \dot{m}^{-1/2} \alpha \; , \label{eq:v0} \\
p_{\rm i} &= (2.3 \times 10^{15} \; {\rm dyn \; cm^{-2}}) \; m^{-1} \dot{m}^{-3/2} \alpha^{-1} \; , \\
T_{\rm i} &= (3.1 \times 10^{7} \; {\rm K})  \; m^{-1/4} \dot{m}^{-3/8} \alpha^{-1/4} \; .
\end{align}
\end{subequations}
In our case, the injection region is near the trapping radius of the disk, which means there will be no photon trapping in the wind so that the wind will stay in the case C1 defined by \citet{Meier1982_PaperII} but never goes to case D. Case C1 assumes that the gas pressure sonic radius ($r_{\rm sg}$), adiabatic radius ($r_{\rm ad}$), and the total pressure sonic radius ($r_{\rm s}$) follow the relation $r_{\rm sg} < r_{\rm ad} < r_{\rm s}$, and the adiabatic region is inside the scattersphere, $r_{\rm ad} < r_{\rm sc}$.  Beyond the injection region, the wind is accelerated by the radiation pressure until the ``critical point'', which is found to be the adiabatic radius ($r_{\rm ad}$) in case C1.

Below the critical radius, the wind develops following the adiabatic and subsonic solutions:
\begin{subequations}
\begin{align}
\rho &= \rho_{\rm i}(r / r_{\rm i})^{-3} \; , \\
v &= v_{\rm i}(r / r_{\rm i}) \; , \\
p &= p_{\rm i}(r / r_{\rm i})^{-4} \; , \\
T &= T_{\rm i}(r / r_{\rm i})^{-1} \; .
\end{align}
\end{subequations}
Above the critical radius, the acceleration stops, and the wind enters into free expansion:
\begin{subequations}
\begin{align}
\rho &= \rho_{\rm i} \zeta^{-1} (r / r_{\rm i})^{-2} \; , \\
v &= v_{\rm i} \zeta \; , \\
p &= p_{\rm i} \zeta^{-1} (r / r_{\rm i})^{-3} \; , \\
T &= T_{\rm i} \zeta^{-1/4} (r / r_{\rm i})^{-3/4} \; ,
\end{align}
\end{subequations}
where $\zeta = r_{\rm ad} / r_{\rm i}$. By solving the following equations under the subsonic condition \citep[see][Equation 14b and the appendix; here $r_{\rm t}$ is the trapping radius and one has $r_{\rm t} \approx r_{\rm i}$ when $\dot{m} \gg 1$]{Meier1982_PaperII},
\begin{align}
r_{\rm ad} &= r_{\rm t} \frac{c_{\rm s}^2}{v^2} \; , \\
c_{\rm s}^2 &= \frac{4p}{3\rho} \; ,
\end{align}
the adiabatic radius, where the flux divergence is comparable to advection, can be determined as
\begin{equation}
r_{\rm ad} = \left(\frac{4p_{\rm i}}{3\rho_{\rm i}v_{\rm i}^2} \right)^{1/4} r_{\rm i} \; .
\end{equation}
The photosphere is the location where the effective absorption optical depth goes to unity, $\tau_\ast = (\tau_{\rm ff/bf} \tau_{\rm es})^{1/2} = 1$, where $\tau_{\rm ff/bf} =  \kappa_{\rm ff/bf} \rho_\ast r_\ast$ is the free-free and bound-free absorption optical depth and $\tau_{\rm es} =  \kappa_{\rm es} \rho_\ast r_\ast$ is the scattering optical depth. In combination with the dynamic structure of the wind and the definitions of the opacities in Table~\ref{tab:def}, the radius, temperature, and scattering optical depth at the photosphere can be readily solved. The subsonic solutions (for $r_\ast < r_{\rm ad}$) are
\begin{align}
r_\ast &= (K_0 \kappa_{\rm es})^{2/7} \rho_{\rm i}^{6/7} T_{\rm i}^{-1} r_{\rm i}^{11/7} \; , \\
T_\ast &= T_{\rm i} (r_\ast / r_{\rm i})^{-1} \label{eq:T_star_sub} \; , \\
\tau_{{\rm es},\ast} &= \kappa_{\rm es}  \rho_\ast  r_\ast =  \kappa_{\rm es}  \rho_{\rm i}  r_\ast^{-2} r_{\rm i}^3  \; ,
\end{align}
and the supersonic solutions (for $r_\ast > r_{\rm ad}$) are
\begin{align}
r_\ast &= (K_0 \kappa_{\rm es})^{8/11} \rho_{\rm i}^{24/11} T_{\rm i}^{-28/11} r_{\rm i}^{27/11} \zeta^{-17/11} \; ,  \\
T_\ast &= T_{\rm i} \zeta^{-1/4} (r_\ast / r_{\rm i})^{-3/4} \label{eq:T_star_sup}  \; , \\
\tau_{{\rm es},\ast} &= \kappa_{\rm es}  \rho_\ast  r_\ast =  \kappa_{\rm es}  \rho_{\rm i} \zeta^{-1}  r_\ast^{-1} r_{\rm i}^2 \; .
\end{align}
An observer at infinity sees the wind photosphere, where the scattering optical depth is still high. Due to radiation transfer, the observed blackbody luminosity reduces to
\begin{equation}
L_{\rm bb} \approx \frac{16 \pi r_\ast^2 \sigma T_\ast^4}{3 \tau_{{\rm es},\ast}} \; .
\label{eq:bb}
\end{equation}
This blackbody luminosity in case C1 is constant since $r_{\rm i}$. By definition, it should be on the order of the Eddington luminosity. Given the boundary conditions defined above, we have $L_{\rm bb} = \frac{3}{4} L_{\rm Edd}$. Equation~\ref{eq:bb} can be used to compare with the data ($T_\ast = T_{\rm bb}$).

Other important radii ($r_{\rm sg}$,  $r_{\rm s}$,  and $r_{\rm sc}$) can be found straightforwardly according to the definitions.  We note that the case C1 assumptions, $r_{\rm sg} < r_{\rm ad} < r_{\rm s}$ and $r_{\rm ad} < r_{\rm sc}$, are satisfied self-consistently for $m \ge 1$ and $10 < \dot{m} < 10^5$.  The choice of $\alpha$ would affect the derived $\dot{m}$. For example, given an observed $T_\ast$ in the subsonic solutions, one gets $\dot{m} \propto \alpha^{2/5}$. This suggests that the derived $\dot{m}$ will be lower by a factor of 2.5 if $\alpha$ goes from 0.1 to 0.01, which is not crucial in this study. 

\end{document}